\documentclass[pdflatex,sn-mathphys]{sn-jnl}% Math and Physical Sciences Reference Style
\jyear{2021}%

\theoremstyle{thmstyleone}%
\theoremstyle{thmstyletwo}%
\newcommand{\Yb}{\mathbf{Y}_{\text{NSE}}}
\usepackage{enumitem}
\theoremstyle{thmstylethree}%

\begin{document}

\title[On Cosmological Low Entropy After the Big Bang]{On Cosmological Low Entropy After the Big Bang: Universal   Expansion and Nucleosynthesis}


\author[1]{\fnm{Charlie F.} \sur{Sharpe}}\email{charliesharpe2101@gmail.com}

\author[2]{\fnm{Luke A.} \sur{Barnes}}\email{l.barnes@westernsydney.edu.au}

\author[1]{\fnm{Geraint F.} \sur{Lewis}}\email{geraint.lewis@sydney.edu.au}

\affil[1]{Sydney Institute for Astronomy, School of Physics, A28, The University of Sydney, NSW 2006, Australia}

\affil[2]{Western Sydney University, Locked Bag 1797, Penrith South, NSW 2751, Australia}


\abstract{
% The Universe’s ability to support complex life is due in part to its expansion during Big Bang nucleosynthesis (BBN) being too rapid for significant fusion of neutrons and protons into heavier elements, leaving a universe that is mostly hydrogen and helium. The small binding energy of these left over nuclides provides the cosmos with a pathway by which entropy can increase, namely through converting nuclear energy into heat and radiation via fusion. This allows for the formation of stars, planets and ultimately life itself. A crucial cosmic parameter in this story is the baryon-to-photon ratio, which results from baryogenesis. Here, we investigate the sensitivity of the Universe's primordial nuclear abundances to changes in both the baryon-to-photon ratio and the temporal evolution of cosmological expansion. We quantify the range of parameter values that produce a substantial change to the degree of fusion during BBN. We find that, while this is indeed linked to baryogenesis and the Universe's expansion history, the requirement of leftover light elements does not place strong constraints on the properties of these two cosmological processes.

%What we do in this paper
We investigate the sensitivity of a universe's nuclear entropy after Big Bang nucleosynthesis (BBN) to variations in both the baryon-to-photon ratio and the temporal evolution of cosmological expansion. Specifically, we construct counterfactual cosmologies to quantify the degree by which these two parameters must vary from those in our Universe before we observe a substantial change in the degree of fusion, and thus nuclear entropy, during BBN.
%
%Why we did it
%We are motivated by Rovelli’s recent claim that a universe’s ability to support complexity, and hence life, relies on its expansion during BBN \cite{Rovelli_2019}. Specifically, the expansion must be rapid enough to prevent significant fusion of neutrons and protons into heavier elements, leaving a universe that consists almost purely of hydrogen and helium. These elements provide a pathway by which nuclear entropy can increase, namely through stellar nuclear fusion.
%
%The small binding energy of these two nuclides provides the cosmos with a post-BBN pathway by which entropy can increase, namely through converting nuclear energy into heat and radiation via fusion. This allows for the formation of stars, planets and ultimately life itself. 
%
%What we found
%We add a crucial cosmic parameter to the story – the baryon-to-photon ratio from baryogenesis. %%%LAB: We said this already
We find that, while the post-BBN nuclear entropy is indeed linked to baryogenesis and the Universe's expansion history, the requirement of leftover light elements does not place strong constraints on the properties of these two cosmological processes.
}




\maketitle

\section{Introduction}\label{sec:intro}
Many physical processes in our Universe, from cell division to star formation, are observed to occur only in one direction in time. The Second Law of Thermodynamics identifies what these processes have in common: they increase the total entropy within a closed, isolated physical system. The Second Law is thus crucial to our understanding of the arrow of time. However, the Second Law is emergent, not fundamental. What light do the fundamental laws of nature shed on the arrow of time? Perhaps surprisingly, with the exception of very rare $T$-violating weak-force processes, these laws show no preference for entropy-increasing over entropy-decreasing processes. Hence, something more is required to explain the Second Law.\footnote{This ``something more'' cannot be mere probability/statistical considerations. These make no reference to time at all. An argument from time-symmetric laws and timeless mathematical principles cannot explain a time-asymmetric universe: the argument could be time-reversed, and be equally valid.} 

The \emph{Past Hypothesis} proposes that this missing piece is a statement about the initial conditions of our Universe: ``the universe had some particular, simple, compact, symmetric, cosmologically sensible, very low-entropy initial macrocondition''\cite{Albert_2015}. With this initial condition, the most likely evolution of our Universe is according to a consistent arrow of time, explaining the success of the Second Law.

However, the Past Hypothesis only postulates the existence of a low-entropy initial macrostate of our Universe. It does \emph{not} specify its form.  So, we can ask: what is it about the arrangement of matter and energy at the start of the Universe that makes it a rare, low-entropy macrostate? Penrose \cite{Penrose_1980,penrose1999} argued that the low entropy of the early Universe is primarily attributed to gravity. The matter is distributed almost uniformly, and the entropy can increase substantially as matter collapses under its own gravity.
We can see this by turning the question around: what would a high-entropy Big Bang look like? Answer: an expanding swarm of black holes, with no energy from gravitational collapse to give.

The central role of gravity has been questioned by Rovelli \cite{Rovelli_2019}, building on Wallace \cite{Wallace_2010}. Most of the entropy-increasing processes we see around us are powered, not by gravitational collapse, but by nuclear fusion. While the gravitational collapse of a gas cloud initially \emph{ignited} the Sun, the last 4.5 billion years of sunlight was powered by the fusion of hydrogen ($^1$H) left over from Big Bang nucleosynthesis (BBN). The Big Bang initial condition is a low-entropy macrostate because it produced a mostly-hydrogen Universe at low temperature. Turning the question around again, a high-entropy Big Bang would look like an expanding gas of iron nuclei, with no energy from nuclear fusion to give.

Rovelli \cite{Rovelli_2019} traces the large abundance of hydrogen in the early Universe to the rapid expansion of space. The Universe, between about one second and three minutes after the Big Bang, expanded too fast for reactions to keep protons and neutrons in equilibrium with each other. This left our Universe in a metastable state, full of cold, diffuse hydrogen. This does not correspond to maximal nuclear entropy\footnote{We use the term ‘nuclear entropy’ when referring to the overall entropy contribution of nuclear fusion.}. The Universe can increase its entropy by burning hydrogen to iron. However, the reaction rate is so slow (outside of stellar cores) that the Universe will remain in this metastable state for a very long time. 
%Gravity, like a struck match, is required only to ignite the reactions. 

The importance of rapid expansion raises the question: just how rapid?\footnote{Rovelli \cite{Rovelli_2019} states that ``the dominant source of the low-entropy of the past universe is only the smallness of the scale factor.'' However, the normalisation of the scale factor in the RW metric is arbitrary, so this condition is not correct. As \cite{Rovelli_2019} states immediately after, it is the fact that the expansion is \emph{rapid} that determines the extent of BBN reactions.} And what cosmological factors determine the critical rate of expansion?
Barnes and Lewis \cite{Barnes_2021} identify a necessary condition for BBN to produce more heavy elements: an excess of baryons over antibaryons in an early universe. Specifically, too much asymmetry increases the baryon-to-photon ratio, allowing nuclear reactions to proceed for longer. By considering the timescales of the relevant nuclear reactions, they conclude that an iron-filled universe requires baryon-antibaryon asymmetry that approaches unity. However, they only consider the standard FLRW ($a \propto t^{1/2}$) expansion of our Universe.

Here, we calculate the effect of altering the early expansion rate of the Universe on BBN and the production of heavy elements. We also vary the amount of baryon-antibaryon asymmetry, via the baryon-to-photon ratio $\eta$. The structure of the paper is as follows. In Section \ref{sec:Background}, we discuss the details of baryogenesis and BBN. In Section \ref{sec:Modelling Modified Universes}, we discuss the different cosmological models that we will consider. In Section \ref{sec:Results}, for each cosmology model, we explain the relationship between post-BBN elemental abundances and variations in both the expansion rate and the baryon-to-photon ratio. We then conclude by discussing these results in Section \ref{sec:Discussion}.

\section{Big Bang Nucleosynthesis}
\label{sec:Background}

Big Bang nucleosynthesis occurs in the era in which protons and neutrons are hot enough to fuse together into deuterium, and the Universe's photon background cool enough that significant amounts of deuterium fuse into heavier nuclei, rather than being photo-disintegrated. In the standard cosmological model, BBN lasted from just under 1 second ($T \sim 2$ MeV) to $10^3$ seconds ($T \sim 0.03$ MeV) after the Big Bang. 

Modelling BBN requires tracking the creation and destruction of nuclear species. The large network of non-linear rate equations is virtually impossible to solve analytically. For this reason, numerical simulations are an attractive option when it comes to modelling BBN, such as those put forward by \cite{Peebles1966a,Peebles1966b,Wagoner1973,Kawano1992,Lisi_1999,Mendoza1999, Pisanti_2008,Arbey_2012,Arbey_2018}. In the following, we use  {\tt AlterBBN}\footnote{{\tt https://alterbbn.hepforge.org/}} \citep{Arbey_2012,Arbey_2018} to compute the final BBN abundances within our counterfactual cosmologies. However, further considerations were required, such as modifying the code (c.f. Appendix \ref{subapp:Modifications To The Code}) and calculating initial abundance conditions (c.f Appendix \ref{subapp:Setting Initial Abundance Coditions}).
Interestingly, the Saha equation can be used to derive analytic expressions for abundances under the condition of Nuclear Statistical Equilibrium (NSE), in which every forward reaction rate is equal to the corresponding reverse reaction rate \citep{Kolb_1990, Barnes_2021}. 

%
%Figure \ref{fig:Standard_cosmo} shows how the abundances of nuclides evolved with time in our Universe.

% \begin{figure}[t!]
%     \centering
%     \includegraphics[width=\textwidth]{Standard_cosmo.pdf}
%     \caption{The temporal evolution of light elements during Big Bang nucleosynthesis in our Universe. We have only plotted nuclides that had an abundance above $10^{-10}$ at some point during BBN with the red dotted line representing a relative abundance of 1. Note that the blue line represents the free neutron abundance as opposed to the total neutron abundance, explaining why the $n/p$ ratio is not constant after NSE breaks.}
%     \label{fig:Standard_cosmo}
% \end{figure}

The baryon-to-photon ratio $\eta$ is particularly important to BBN, as it fixes the relationship between temperature and baryon density; nuclear reaction rates depend on both. This ratio is determined by the baryon-antibaryon asymmetry that the early universe possesses. After baryons and antibaryons annihilated with each other at a temperature around $2 \times 10^{12}$ K \citep{liddle_2015}, our Universe only consisted of photons and leftover baryons. Today, we measure $\eta_0 = 6.1 \times 10^{-10}$, indicating that for (roughly) every one billion anti-baryons, there were one billion and one baryons before annihilation \citep{Bennett2003}. The degree of baryon-antibaryon asymmetry is determined by the process of baryogenesis, which is currently not well understood.

In the counterfactual cosmological models we consider in this paper, we find that fusion continues down to lower temperatures. As a result, we model down to temperatures much lower than 0.03 MeV. The initial and final temperatures we consider are $T_i \sim 2.7$ MeV and $T_f \sim 0.001$ MeV, respectively \citep{Kolb_1990}. Due to the temperature at which fusion occurs being independent of a universe's expansion rate and baryon-to-photon ratio, we keep these values of $T_i$ and $T_f$ fixed, independent of the cosmological model being considered.

\section{Modelling Modified Universes}
\label{sec:Modelling Modified Universes}

Here, we discuss below our strategy for altering the expansion rate, which we achieve by altering the function $a(t)$. We will then be in a position to answer our question: how rapid does a universe's expansion need to be for significant amounts of nuclei to be fused into heavy elements during BBN?
The point here is \emph{not} to attempt to model the actual history of our Universe. We are exploring the physical relationship between the expansion of a universe and the low-entropy nuclear energy available post-BBN. We are not proposing alternative models of the early Universe, or exotic new forms of energy.

\subsection{The Forced Cosmology}
\label{subsubsec:The Forced Cosmology}
In this model, we specify the scale factor's dependence on time by fiat: $a(t) \propto t^n$. We label this ‘the forced cosmology’. The energy density of matter and radiation depend on the scale factor in the usual way ($a^{-3}$ and $a^{-4}$, respectively).

\begin{figure}[t!]
    \centering
    \includegraphics[width=\textwidth]{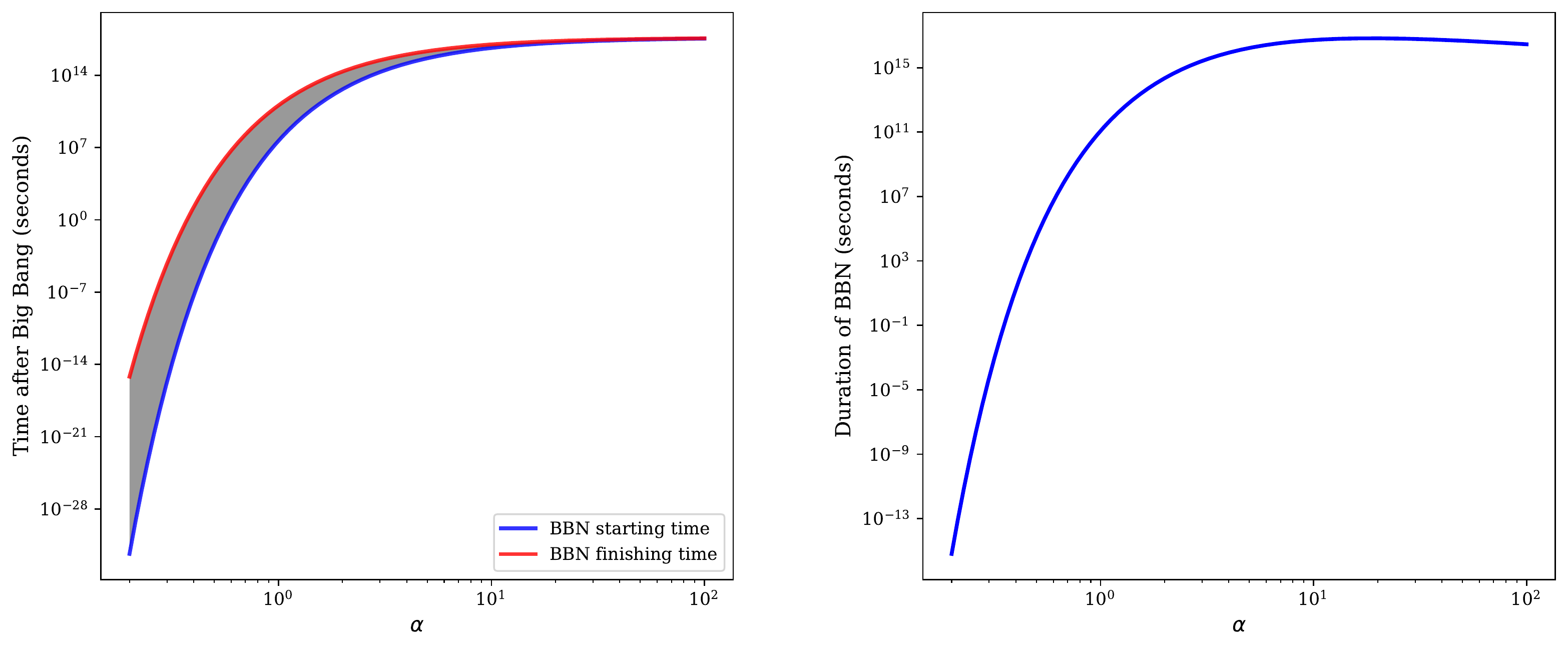}
    \caption{Left Panel: This shows how the starting and finishing time of BBN change with $\alpha$ for the forced cosmology. From Equation \ref{eq:t(T) for forced}, the variation of temperature over the BBN periods can be easily understood from the fact that an order of magnitude increase in time corresponds to temperature dropping by $\alpha$ orders of magnitude. Right panel: This shows how the duration of BBN varies with $\alpha$. While this can be derived from the left panel, the right panel provides easier visualisation. Here, $\alpha$ only goes down to 0.2 for reasons discussed in Section \ref{subsec:Results - the Forced Cosmology}. Note that because we have normalised the power law to ``today'' ($t_0$ and $T_0$), for the range of $\alpha$ that we consider here, larger values of $\alpha$ imply a \emph{longer} duration of BBN. This is because they reach the relevant BBN temperatures later in their evolution.}
    \label{fig:Time_of_BBN_for_forced_uni}
\end{figure}

The temperature of the cosmic microwave background is inversely proportional to the scale factor $T \propto a^{-1}$, and $a \propto t^\alpha$. This gives
\begin{equation}
    t = t_0 \left(\frac{T_0}{T}\right)^{1/\alpha} ~, 
    \label{eq:t(T) for forced}
\end{equation}
where we set $T = T_0$ at $t = t_0$, where $T_0 = 2.725$ K is the temperature of CMB today and $t_0 = 13.7$ Gyr is the age of our Universe today.

The starting time of BBN is found by substituting $T = T_i$. Having $a \propto T^{-1}$ and $a \propto t^{\alpha}$ gives the relationship between the finishing time of BBN, $t_f$, and the starting time of BBN, $t_i$, to be $t_f = (T_i/T_f)^{1/\alpha}t_i$. For the rest of this paper, we will define the duration of BBN to be the time required for the temperature to go from $T_i$ to $T_f$, despite element abundances mostly freezing out at earlier temperatures. Thus, using the approximate values of the BBN starting and finishing temperatures that are given in \cite{Arbey_2012}, namely $T_i = 2.7 \times 10^{10}$ K ($2.32$ MeV) and $T_f = 10^7$ K ($8.6 \times 10^{-4}$ MeV), we are able to examine the period over which BBN occurs as a function of $\alpha$. This is shown in Figure \ref{fig:Time_of_BBN_for_forced_uni}. We can see that the duration covers many orders of magnitude.
Note that because we have normalised the power law to ``today'' ($t_0$ and $T_0$), larger values of $\alpha$ imply a \emph{longer} duration of BBN. This is because they reach the relevant BBN temperatures later in their evolution.

\subsection{The Dominant Fluid Cosmology}
\label{subsubsec:The Dominant fluid Cosmology}

% In this model, we change the expansion rate by adding a new form of energy and calculating the effect on the expansion rate. The evolution of the scale factor and energy densities are linked by the Friedmann equation and the fluid equation,
% \begin{align}
%     \left(\frac{\dot{a}}{a}\right)^2 = \frac{8\pi G\rho}{3} - \frac{kc^2}{a^2} + \frac{\Lambda}{3}, \label{eq:Friedmann eqn}\\
%     \dot{\rho} + 3\frac{\dot{a}}{a}\left(\rho + \frac{P}{c^2}\right) = 0 ~, \label{eq:Fluid eqn}
% \end{align}
% where $G$ is the gravitational constant, $\rho$ is the energy density of the universe, $k$ is the curvature of the universe, $c$ is the speed of light, $\Lambda$ is the cosmological constant, and $P$ is the pressure density of the universe.

The dominant fluid cosmology adds to the standard model of cosmology a form of energy that dominates during the period of Big Bang nucleosynthesis, whose energy density we label $\rho_D$. By requiring consistency with the Friedmann equation and fluid equation, this additional dominant fluid will control the rate of expansion. The equation of state (EoS) of the dominant fluid is given by $w = P / (c^2 \rho_D)$, and for simplicity is assumed to be constant throughout BBN.

Upon solving the Friedmann equation and fluid equation, one finds $\rho_D \propto T^{3(1+w)}$ and $a \propto t^{2/(3w+3)}$ (assuming $\rho_D$ dominates over all other energy forms) \cite{Kolb_1990}. The difference between this model and the forced cosmology lies in the determination of the initial conditions, and duration, of BBN. Hereafter, we will be working in natural units ($c = \hbar = k_B = 1$) unless stated otherwise.

The time independence of the EoS parameter restricts the scale factor to a power-law \cite{Kolb_1990}, $a \propto t^{\alpha}$, giving the EoS parameter of the dominant fluid to be $w = 2/(3\alpha) - 1$. Thus, $\rho_D = AT^{2/\alpha}$ for some constant $A$. We find that the starting time of BBN is given by,
\begin{equation}
    t_i = 
    \begin{cases}
    \dfrac{3\alpha T_i^{-1/\alpha}}{\sqrt{24\pi G A}} & \ \quad \text{if } \alpha \leq 1/2, \vspace{0.15cm}\\
    \dfrac{T_i^{-1/\alpha}}{\sqrt{\frac{32}{3}\pi G A}} & \ \quad \text{if } \alpha > 1/2,
    \end{cases} \label{eq:dom fluid starting time}
\end{equation}
where 
\begin{equation}
    A = 
    \begin{cases}
    \text{max}\left[\dfrac{43\pi^2}{120}(T_f)^{4-2/\alpha},\dfrac{2 \pi^2 m_p \eta}{81}(T_f)^{3-2/\alpha}\right] & \quad \text{ if } \alpha \leq \frac{1}{2} \vspace{0.13cm},
    \\
    \text{max}\left[\dfrac{43\pi^2}{120}(T_i)^{4-2/\alpha},\dfrac{2 \pi^2 m_p \eta}{81}(T_f)^{3-2/\alpha}\right] & \quad \text{ if } \frac{1}{2} < \alpha \leq \frac{2}{3} \vspace{0.13cm},
    \\
    \text{max}\left[\dfrac{43\pi^2}{120}(T_i)^{4-2/\alpha},\dfrac{2 \pi^2 m_p \eta}{81}(T_i)^{3-2/\alpha}\right] & \quad \text{ if } \alpha > \frac{2}{3}.
    \end{cases}
    \label{eq:app_A_min}
\end{equation}
with $m_p$ being the mass of the proton. We note that the expression for $A$ gives the minimum value required for $\rho_D$ to dominate during BBN. While $A$ could be larger, we made this assumption to remove ambiguity when determining the times and temperatures at which dominance switches between $\rho_m,\rho_r$ and $\rho_D$.
\begin{figure}[t!]
    \centering
    \includegraphics[width=\textwidth]{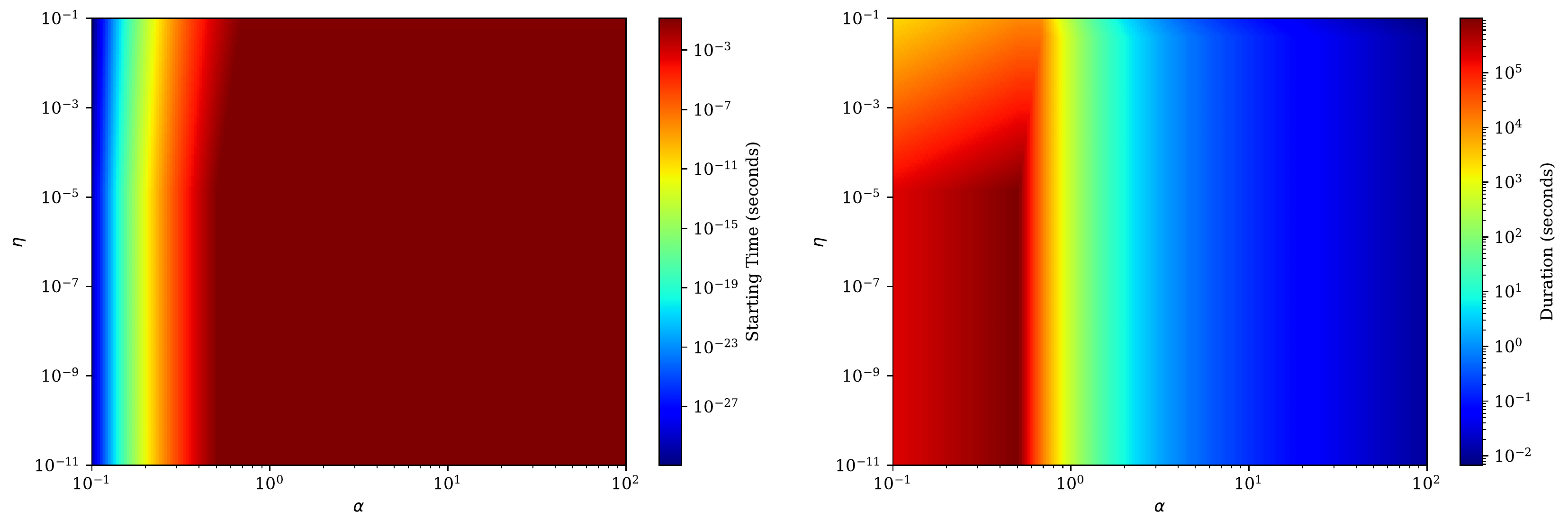}
    \caption{Left Panel: This shows how the starting time of BBN changes with $\alpha$ and $\eta$ for the dominant fluid cosmology. Right panel: This shows how the duration of BBN varies with $\alpha$ and $\eta$.}
    \label{fig:Time_of_BBN_for_dominant}
\end{figure}

Figure \ref{fig:Time_of_BBN_for_dominant} shows how the duration, $t_f - t_i$, of BBN varies with $\alpha$ and $\eta$. The range of BBN durations in this cosmological model is much smaller than that in the forced cosmology. It should also be noted that the forced cosmology can be derived from the dominant fluid cosmology by forcing $t_0 = T_0$ and taking $A \to \infty$.

\section{Results}
\label{sec:Results}

Here, we present how the final nuclide abundances were effected by variations in $\alpha$ and $\eta$, calculated using our modified version of {\tt AlterBBN}. The easiest way to do so is by studying the most abundant, and second most abundant, nuclides left over after BBN. We will discuss the forced cosmology and the dominant fluid cosmology separately. Altogether, this paper reports on the results from $11680$ BBN simulations.

\subsection{The Forced Cosmology}
\label{subsec:Results - the Forced Cosmology}

\begin{figure*}[t!]
    \centering
    \includegraphics[width=\linewidth]{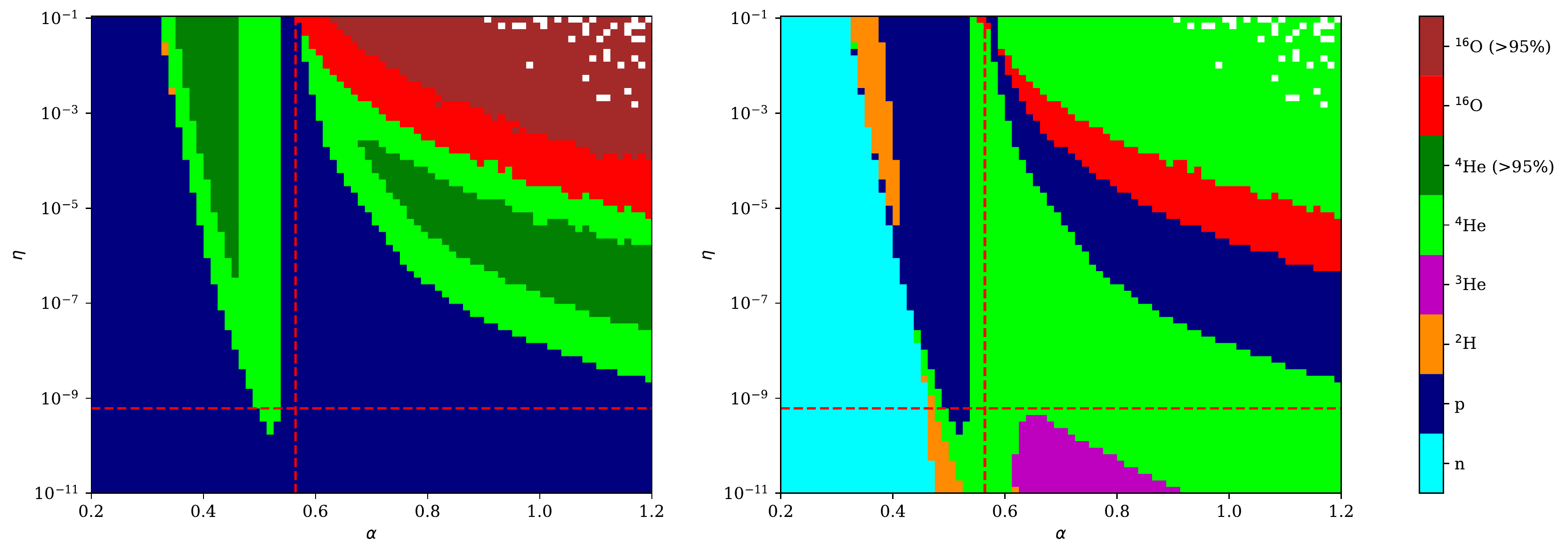}
    \caption{The most abundance (left panel), and second most abundance (right panel), nuclides left over from BBN in the forced cosmology as a function of $\alpha$ and $\eta$. The horizontal red dashed lines represent the value of $\eta$ in our Universe while the vertical ones represent the value of $\alpha$ required for BBN duration to match that of our Universe. In the left panel, we have split the regions dominated by $^{4}$He and $^{16}$O into those where the abundances are greater than 0.95 and less than 0.95.  Here, we have varied $\alpha$ from $0.2$ to $1.2$, and $\eta$ over 10 orders of magnitude, from $10^{-11}$ to $10^{-1}$. The white regions represent the parameters where the code was unable to complete the calculation due to the convergence issues. Note that the jagged boundaries are due to our use of a discrete set of simulation parameters.}
    \label{fig:A0_1_1st_2nd_dom_comparison}
\end{figure*}

Figure \ref{fig:A0_1_1st_2nd_dom_comparison} shows how the most abundant, and second most abundant, nuclides left over from BBN vary with $\alpha$ and $\eta$ in the forced cosmology. We have ignored $\alpha < 0.2$ as BBN duration is $\mathcal{O}(10^{-15}\text{s})$ and hence $^1$H will always dominate as there is not enough time for any fusion to occur. Convergence issues within the code forced us to set $\alpha < 1.2$, which we suspect was due to the large time-steps leading to invalid linear approximations within the {\tt AlterBBN} (see Ref. \cite{Arbey_2018} for details of the linearization process). Updating the integration methods as a means of resolving these issues is beyond the scope of this paper. We also only consider $\eta \leq 10^{-1}$ due to {\tt AlterBBN} not considering degeneracy in it's calculations.

The complicated trend in Figure \ref{fig:A0_1_1st_2nd_dom_comparison} stems from the intricate interplay between $\alpha$, $\eta$, the starting time of BBN, the duration of BBN, and the NSE configuration when NSE breaks. This last detail is most important.

Let $\Yb(T)$ be the NSE abundance configuration at temperature $T$. As temperature decreases, heavier elements will take over as the dominant species. Let $T_k$ be the temperature at which elemental species $k$ first becomes dominant. Over the temperature range of BBN, the dominating NSE element transitions from $^1$H$ \to ^4$He$ \to ^{16}$O with $T_{^1 \text{H}} > T_{^4 \text{He}} > T_{^{16} \text{O}}$. It is the form of $\Yb(T)\vert_{T = T_{\text{NSE}}}$, where $T_{\text{NSE}}$ is the temperature at which NSE breaks, that is ultimately responsible for the observed trend. Note $T_k$ is independent of $\alpha$ and monotonically increases with $\eta$ \cite{Kolb_1990}, while $T_{\text{NSE}}$ is independent of $\eta$ and monotonically decreases with $\alpha$ (c.f. Appendix \ref{subapp:Setting Initial Abundance Coditions}).

When $T_{^1\text{H}} > T_{\text{NSE}} > T_{^4\text{He}}$, which is the case for $\alpha \lesssim 0.54$, the neutron to proton ratio will freeze out at a value of $X_n/X_p = \exp(Q/T_{\text{NSE}})$ where $Q = 1.29$ MeV is the mass difference of a proton and neutron. This creates a cap on the possible $^4$He production. The production will get closer to this cap as $\eta$ increases due to a larger portion of the free neutrons being fused in $^4$He nuclei.

When $\alpha < 0.5$, $X_n/X_p \approx 1$ since $T_{\text{NSE}} \gg Q$, leading to an almost absent cap. Hence, depending on $\eta$, we can effectively reach arbitrarily strong $^4$He dominance. However, once $\alpha \approx 0.54$, $X_n/X_p$ falls below 0.5, capping the maximum $^4$He abundance at a value below that of $^1$H, returning us to a $^1$H dominated universe. This is then made worse by free neutron decay. 

A further increase in $\alpha$ eventually gives $T_{^4\text{He}} > T_{\text{NSE}} > T_{^{16}\text{O}}$ and we hence arrive back at $^4$He dominance where the lack of $^{16}$O production is due to insufficient time for fusion into $^{16}$O. Increasing $\alpha$ even further leads to $T_{^{16}\text{O}} > T_{\text{NSE}}$, and we get strong $^{16}$O dominance.

As we increase $\eta$ for $\alpha < 0.54$, we get closer to the $^4$He cap, thus leading to an increase in $^4$He production. Furthermore, as $\eta$ increases, $T_{^1 \text{H}}, T_{^4 \text{He}}, T_{^{16} \text{O}}$ also increase while $T_{\text{NSE}}$ remains the same. Thus, on the right, we expect a negative gradient for the $^4$He and $^{16}$O islands, which is what we observe. 

Additionally, $^{16}$O is the heaviest element that {\tt AlterBBN} considers, and hence we are unable to quantitatively conclude whether fusion in universes that are $^{16}$O dominated would continue all the way up to iron. However, it is unlikely that the fusion would stop at $^{16}$O.

\subsection{The Dominant Fluid Cosmology}
\label{subsec:Results - the Dominant Fluid Cosmology}

\begin{figure*}[t!]
    \centering
    \includegraphics[width=\linewidth]{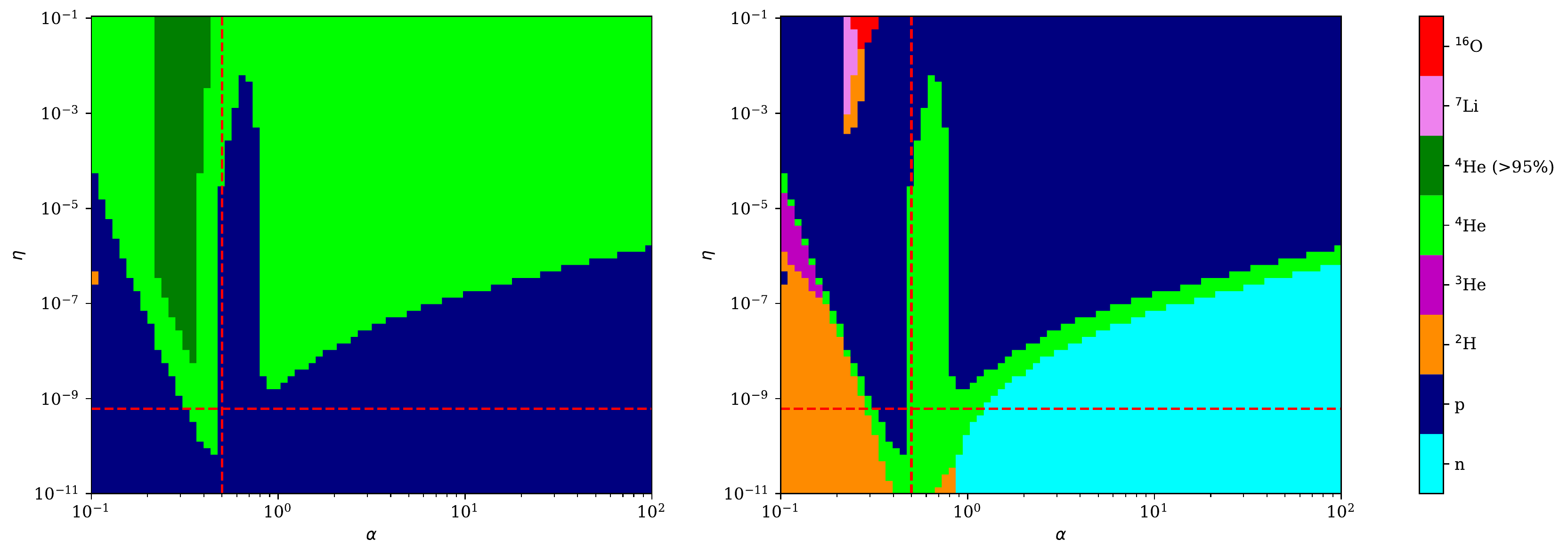}
    \caption{The most abundance (left panel), and second most abundance (right panel), nuclides left over from BBN in the dominant fluid cosmology as a function of $\alpha$ and $\eta$. The horizontal red dashed lines represent the value of $\eta$ in our Universe while the vertical ones represent the value of $\alpha$ required for BBN duration to match that of our Universe. In the left panel, we have split the regions dominated by $^4$He into those where the abundance is greater than 0.95 and less than 0.95. Here, we have varied $\alpha$ over 3 orders of magnitude, from $10^{-1}$ to $10^2$, and $\eta$ over 10 orders of magnitude, from $10^{-11}$ to $10^{-1}$. Again, note that the jagged boundaries are due to our use of a discrete set of simulation parameters.}
    \label{fig:Amin_1st_2nd_dom_comparison}
\end{figure*}

Figure \ref{fig:Amin_1st_2nd_dom_comparison} shows which two nuclides are most abundant in the dominant fluid cosmology for varying values of $\alpha$ and $\eta$. We vary $\alpha$ from $10^{-1}$ to $10^2$ and $\eta$ from $10^{-11}$ to $10^{-1}$ and have no convergence issues. We did not consider values of $\eta > 10^{-1}$ for the same reasons as in the forced cosmology.

We first note that universes with $\alpha < 0.2$ do not experience NSE before BBN started (c.f. Appendix \ref{subapp:Setting Initial Abundance Coditions}). Hence, all we are allowed to note here is the lack of significant fusion, despite choosing an initial configuration identical to our Universe.
We find that $T_{^1\text{H}} > T_{\text{NSE}} > T_{^4\text{He}}$ for the entire parameter space considered. The behaviour of the abundances for $0.2 < \alpha < 0.5$ is almost identical to the forced cosmology. We, again, obtain almost arbitrarily strong $^4$He dominance for $0.2 < \alpha < 0.4$. We also observe that $X_n/X_p$ falls below 0.5 at $\alpha \approx 0.48$, returning us to a $^1$H dominated universe. Interestingly, in the dominant fluid cosmology, $T_{\text{NSE}}$ increases with $\eta$ for $\eta \gtrsim 2 \times 10^{-5}$ (c.f. Appendix \ref{subapp:Setting Initial Abundance Coditions}). For $\eta \gtrsim 2 \times 10^{-3}$, this increase is great enough to prevent $X_n/X_p$ from dropping below 0.5 for any values of $\alpha$ and hence we do not return to $^1$H dominance.

For $\alpha > 0.5$, unlike in the forced cosmology, we find that $T_{\text{NSE}}$ begins to increase, asymptotically approaching $T_i$. Thus, $X_n/X_p$ increases with $\alpha$ and we go back to $^4$He dominance. Finally, as we take $\alpha$ to be very large, we will eventually return back to $^1$H dominance as the duration of BBN will become very small, explaining the positive gradient on the bottom boundary of the $^4$He island on the right.

\section{Discussion}
\label{sec:Discussion}

Our calculations assume that no relativistic species are degenerate, quantified by $\mu_i \ll T$ where $\mu_i$ is chemical potential of the species, $i$, in consideration. If this criterion is not met, then we are unable to assume $\rho_r \propto T^4$. The most important relativistic species during BBN is the electron, which \cite{mukhanov_2005} (pg. 93) states to obey $\mu_e/T \sim B$, where $B$ is the baryon-to-entropy ratio. Furthermore, we know that $B \sim \eta$ and hence non-degeneracy can be assumed if $B \sim \eta \ll 1$ \cite{mukhanov_2005, Barnes_2021}. The largest value of $\eta$ we consider is $10^{-1}$, validating our assumption.

As we slow down the rate of expansion during BBN, while keeping $\eta = 6.1 \times 10^{-10}$ constant, Rovelli suggests we should see the dominant nuclides change from light elements to heavy elements. This trend is eventually observed within the forced cosmology when we reach large $\alpha$. However, the trend is not monotonic. The universe with the greatest heavy element production occurred when $\alpha \approx 0.50$, where the abundances of $^1$H and $^4$He were 0.329 and 0.634, respectively. Almost all other values of $\alpha$ gave universes that were $^1$H dominated, as can be seen in Figure \ref{fig:A0_1_1st_2nd_dom_comparison}. This corresponds to a very short BBN duration of about 9 hours, as opposed to about 13 days in our Universe (we remind the reader that we have defined the duration to be the time taken for $T$ to drop from $T_i$ to $T_f$, rather than when the abundances freeze out).

For the dominant fluid cosmology, we did not observe the trend suggested by Rovelli. We found that the universe with the greatest heavy element production occurred when $\alpha \approx 0.37$, where the abundances of $^1$H and $^4$He were 0.169 and 0.818, respectively. This corresponds to a BBN duration about 20\% shorter than in our Universe. We do not expect values of $\alpha$ greater than those considered here, i.e. $\alpha > 10^2$, to provide heavy element build up due to the duration of BBN decreasing and the starting time remaining the same as $\alpha$ increases. While the aforementioned universe is $^4$He dominated, with $^4$He providing less access to nuclear energy than $^1$H, it still possesses low nuclear entropy and has a substantial amount of $^1$H left over. Again, almost all other values of $\alpha$ gave universes that were $^1$H dominated. This agrees with results presented by \cite{Barnes_2021} which show that BBN in our Universe would have to last approximately 1 billion years for all elements to burn all the way to iron. 

Note that $^{16}$O is the heaviest element that {\tt AlterBBN} considers, so we don't predict whether fusion all the way up to iron would have occurred. However, the trend towards heavier elements is clear. Approximate calculations of fusion up to iron can be found in \cite{Barnes_2021}.

When allowing variations in $\eta$, we that see a large portion of the parameter space of the forced cosmology has strong $^4$He dominance (greater than 95\% $^4$He), which represents universes that are far from maximal nuclear entropy, but still closer than our own. This occurs for $0.335 < \alpha < 0.475$ and $10^{-6} < \eta < 10^{-1}$. These values of $\alpha$ give a duration of BBN between about 16 seconds and 2 hours, much shorter than in our Universe, and require a value of $\eta$ multiple orders of magnitude larger than $\eta_0$. We also see strong $^4$He build up in the central part of the right $^4$He island. Furthermore, as we move above this region, $^4$He is fused to $^{16}$O and hence these universes move even closer to maximal entropy. Note, however, that the BBN duration in these universes is around 31 years and they require a value of $\eta$ many orders of magnitude greater than $\eta_0$.

In the dominant fluid cosmology, the maximum abundance that $^{16}$O reaches is around 13\%, which occurs when $\eta$ is more than 8 orders of magnitude larger than $\eta_0$ and a duration around 50 times shorter than in our Universe. This lack of heavy element build up makes sense when considering that BBN duration only varies between $10^{-2} - 10^6$ seconds in this cosmology. We see a large section of strong $^4$He dominance within the approximate parameter range of $0.2 \lesssim \alpha \lesssim 0.4$ and $10^{-8} \lesssim \eta \lesssim 10^{-1}$. This corresponds to a duration of BBN to be slightly smaller than our Universe, between about 5 days ($\sim 434,000$ seconds) and 9 days ($\sim 782,000$ seconds), and requires $\eta$ to be at least more than 2 orders of magnitude greater than its current value. While we acknowledge these universes have greater nuclear entropy than ours, we emphasise that this range is very narrow when considering all the permissible values of $\eta$ and BBN duration.

In conclusion, while an extremely slow cosmological expansion would have led to considerable heavy element production, the duration of BBN can be varied by many orders of magnitude without resulting in a universe that burns entirely to heavy elements. To put it another way, BBN in our Universe is many orders of magnitude more rapid than is required for a low-nuclear-entropy initial conditions. Further, the relationship between time available for BBN and element abundance is far from monotonic. The answer to the question ``how rapid?'' seems to be ``faster than a billion years''.

\section{Acknowledgments}
The majority of the work presented here was undertaken as part of C. Sharpe's honours year at the University of Sydney, but was originally conceived by G. F. Lewis. We would like to thank A. Arbey et al. for making their {\tt AlterBBN} code publicly available as this paper would not have been possible without it. G. F. Lewis received no funding to support this research. C. Sharpe would also like to thank NSW police for ensuring a swift return of his belongings, including his laptop, after having them burgled from his house a few weeks before this paper's submission. Additionally, he would like to thank his neighbour, Gary, who spotted the burglar and, rather than simply phoning the police and staying put, decided to yell `you better run fast mate' before chasing the man down the street, tackling him, pinning him to the ground, and \textit{then} calling the police, all while still in his pyjamas and a sleepy daze.

\subsection*{Declarations}

\textbf{Competing interests:} None

\begin{appendix}

\section{AlterBBN}
\label{app:AlterBBN}

\subsection{Modifications to the Code}
\label{subapp:Modifications To The Code}

In their code, Arbey et al. have an incorrect expression for the starting time which is many orders of magnitude smaller than it should be. This was pointed out by \cite{Sharpe_2021}. Despite this, the current publicly available version of the code does not run into issues due to the abundances and temperature being forced to not change until $t = 0.136$ seconds $-$ the starting time of BBN within our Universe $-$ is reached. When varying $\alpha$ and $\eta$, we had to remove this condition and correct the starting time. 

Furthermore, the reaction rate of three reactions, $^2$H$ + p \to \gamma + ^3$He,  $^2$H$+ ^2$H $\to n + ^3$He and $^2$H$+ ^2$H $\to p + ^3$He, obey step functions. This produced unphysical jigsaw patterns in the time evolution of the abundances, and so we replaced these expressions with a linear interpolation of the discrete values\footnote{This was merely to make interpretation easier, the results of our calculations are unaffected.}. 

Finally, the fluid equation is used to calculate the rate of change of $\ln(a^3)$ with respect to temperature, which is then used to calculate $dT/dt$. This would occasionally cause $dT/dt \geq 0$ despite $da/dt > 0$, which is clearly unphysical. Hence, using the relation $a \propto T^{-1}$, we replaced their calculation of $d \ln(a^3)/dT$ with $d \ln(a^3)/dT = -3/T$, ensuring that $dT/dt < 0$ at all times. 

\subsection{Setting Initial Abundance Conditions}
\label{subapp:Setting Initial Abundance Coditions}

By default, {\tt AlterBBN} determines the starting abundances for neutrons, protons, and deuterium by their NSE configuration at $T = T_i$ (c.f. Ref. \cite{Kolb_1990} p.88). 
This is only valid if $T_{\text{NSE}} < T_i$. NSE is broken when any reaction rate falls below $H$. The first to succumb is always $\Gamma_{pe \to \nu n}$ and hence by computing the temperature, $T_{\text{NSE}}$, at which $\Gamma_{pe \to \nu n} = H$, we are able to determine when NSE is broken and thus whether we can assume an NSE initial abundance configuration.
We achieve this by using the definition of the Hubble parameter, $H = \dot{a}/a = \alpha/t$, Equations \ref{eq:t(T) for forced} and \ref{eq:dom fluid starting time}, and the relatioship
\begin{equation}
    \Gamma_{pe \to \nu n} = \frac{7}{60} \pi\left(1+3 g_{A}^{2}\right) G_{F}^{2} T^{5}. \label{appeq:n->p reaction rate}
\end{equation}
Here, $g_A$ is the axial-vector coupling of the nucleon\footnote{With numerical value of 1.26.} and $G_F$ is the Fermi coupling constant\footnote{With numerical value $1.16 \times 10^{-5}$ GeV$^{-2}$.} \citep{Kolb_1990}. Note that this equation is only valid in the high temperature limit, $T \gg 1.28$ MeV. However, this does not concern us as we are only interested in whether NSE is achieved at $T = T_i = 2.32$ MeV. 

\begin{figure}[t!]
    \centering
    \includegraphics[width=\textwidth]{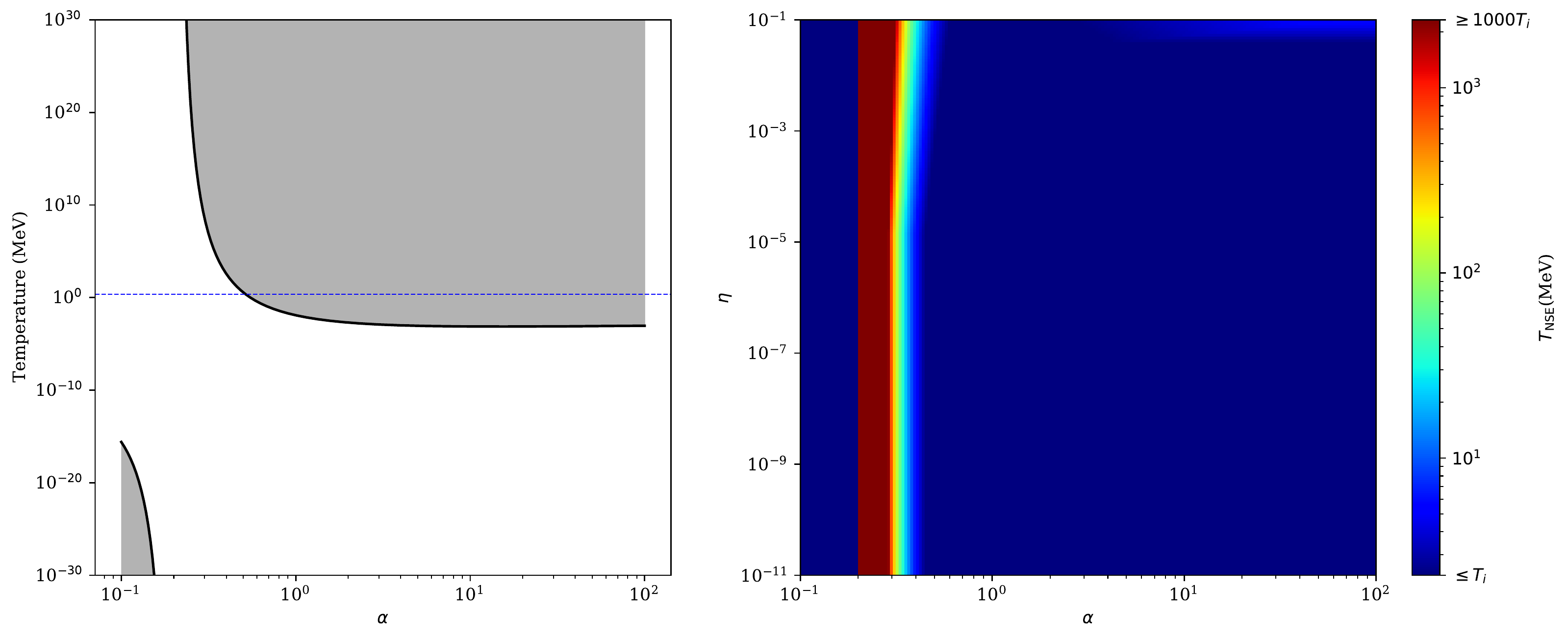}
    \caption{Left panel: The temperature range, shown in grey, over which $\Gamma_{pe \to \nu n}/H \geq 1$ for the forced cosmology with the blue line representing $T = T_i$. We only have dependence on $\alpha$ here. Right panel: The temperature at which $\Gamma_{pe \to \nu n}/H = 1$ as a function of both $\alpha$ and $\eta$. To the left of $\alpha < 0.2$, the universe moves \textit{into} NSE like in the forced cosmology.}
    \label{fig:NSE_break_temp_vs_alphs}
\end{figure}

The relationship between time and temperature is independent of $\eta$ for the forced cosmology, while this is not the case for the dominant fluid cosmology. Accordingly, in Figure \ref{fig:NSE_break_temp_vs_alphs}, we have shown $T_{\text{NSE}}$ against $\alpha$ only for the forced cosmology (left panel) and $T_{\text{NSE}}$ against both $\alpha$ and $\eta$ for the dominant fluid cosmology (right panel).
Since $\Gamma_{pe \to \nu n}/H \propto T^{(5\alpha-1)/\alpha}$, this ratio increases with time for $\alpha < 0.2$. Additionally, $\Gamma_{pe \to \nu n}/H \ll 1$ at $T = T_i$. This indicates that these universes do not experience a period of NSE whatsoever before BBN starts. This makes it troublesome to choose an initial configuration as knowledge of the conditions of the much earlier Universe is required, which we do not have. For the purposes of this report, and due to the lack of a better choice, we will assume the initial abundances for these universes are the same as that of BBN in our Universe.

We also acknowledge that the quark-gluon plasma condensed into nucleons when $T \sim 1$ GeV \citep{choudhuri_2010}. Thus, when $T_{\text{NSE}} \gtrsim 1$ GeV, we simply set $X_n/X_p = \exp(-Q/T_{\text{NSE}})$ as further considerations are beyond the scope of the paper.

We see that $T_{\text{NSE}} > T_i$ for $0.2 < \alpha \leq 0.55$ in the forced cosmology, and $0.2 < \alpha \lesssim 0.45$ in the dominant fluid cosmology. Thus, we set the initial proton and neutron abundances to be those at the time NSE was broken. This is valid as the time between $T_{\text{NSE}}$ and $T_i$ for these values of $\alpha$ is less than 0.2 seconds and so significant neutron decay won't not occur.

Deuterium still follows its NSE abundance at $T = T_i$ for $\alpha \geq 0.3$ in the forced cosmology and $\alpha \geq 0.16$ in the dominant fluid cosmology. These were both checked numerically with {\tt AlterBBN}. For $\alpha < 0.3$ in the forced cosmology, BBN is too rapid for any fusion to occur between $T_{\text{NSE}}$ and $T_f$ and hence the post-BBN configuration consists purely of protons and neutrons. Recall that for $\alpha < 0.2$, we are assuming an NSE configuration at $T_i$, as discussed above, and hence no further considerations are required.

\subsection*{Data Availability}

The datasets generated and analysed during the research carried out in this paper are available from C.S. on reasonable request.




\end{appendix}


\bibliography{sn-bibliography}% common bib file
%% \input sn-article.bbl


\end{document}